\documentclass[11pt,reqno]{article}
\usepackage{geometry}                
\usepackage{graphicx}
\usepackage{amssymb}
\usepackage{authblk}
\usepackage{epstopdf}
\graphicspath{{./figs/}}
\begin{document}

\title{Solution of the Generalized Linear Boltzmann Equation for Transport in Multidimensional Stochastic Media}
\author{Ari Frankel}
\affil{Sandia National Laboratories, P.O. Box 969, Livermore, CA 94551}

\maketitle

\begin{abstract}
The generalized linear Boltzmann equation (GLBE) is a recently developed framework based on non-classical transport theory for modeling the expected value of particle flux in an arbitrary stochastic medium. Provided with a non-classical cross-section for a given statistical description of a medium, any transport problem in that medium may be solved. Previous work has only considered one-dimensional media without finite boundary conditions and discrete binary mixtures of materials. In this work the solution approach for the GLBE in multidimensional media with finite boundaries is outlined. The discrete ordinates method with an implicit discretization of the pathlength variable is used to leverage sweeping methods for the transport operator. In addition, several convenient approximations for non-classical cross-sections are introduced. The solution approach is verified against random realizations of a Gaussian process medium in a square enclosure.
\end{abstract}

\section{Introduction}

Stochasticity in the spatial distribution of participating media can have a large impact on the statistics of particle transport. Pebble-bed nuclear reactor neutron fluxes are modified by the distribution of the fuel pebbles compared to predictions from an equivalent homogenization \cite{LarsenVasques11}. In high temperature combustion, turbulence can induce highly non-uniform gas densities and blackbody radiation intensities, leading to enhanced thermal emissions from flames \cite{TRI1,TRI2}. The stochastic distribution and structure of clouds in the atmosphere have various impacts on climate models of solar and terrestrial radiation \cite{MarshakDavis}. The transmission of solar energy through particle curtains subjected to turbulence is also of concern for solar energy devices \cite{Sandia}.\\

Transmission through a spatially correlated medium has been studied extensively in recent years. It was shown in \cite{Shaw02,Kostinski02,Borovoi,KostinskiReply} that depending on whether the medium is positively or negatively correlated at the microscopic level, radiation may undergo sub-exponential or super-exponential attenuation respectively relative to the rate predicted from Beer's law. In \cite{MarshakDavis04}, the spectra of different stochastic processes were analyzed and related to observed rates of radiation attenuation. \\

Due to the deviations from the classic transport predictions, as well as the possible variance in results, it is important to develop methods to capture this variability. Monte Carlo sampling of random media is an unbiased way of gathering this information, but this requires many samples of the media which may not be available, as well as a solution of the transport equation in each sample. Such studies can become prohibitively costly. In some cases, it is possible to perform dimension reduction \cite{Olson17,Fichtl11} to reduce the number of samples required, but this is not always guaranteed and requires an identification of the spectrum of the stochastic medium. In the limit of short length-scales and mild variability, the so-called atomic mixing approximation \cite{AtomicMix} may be valid and allow one to neglect stochastic effects, but this limit is prohibitive and inaccurate for many cases of interest. Alternatively, one might seek an effectively reduced cross-section that captures the variability \cite{Shaw02,Frankel17}, but this approximation is only valid for mean-free-paths much longer than the length scale of variability. A commonly used approach in mixtures of discrete binary media is the Levermore-Pomraning model \cite{LevermorePomraning}. With some approximations in the nature of the transport through the medium, a coupled set of transport equations for the average angular flux may be derived and solved using existing transport methods. Though this model is convenient, it suffers from a loss of accuracy at higher scattering ratios and is obviously restricted to the special case of binary Markovian media.\\

A recently developed model called the Generalized Linear Boltzmann Equation (GLBE) is related to non-classical particle transport, in which the particles are considered to have a probability of extinction that depends on how far they have already traveled \cite{LarsenVasques11}. This model unifies all possible forms of stochastic media into a non-classical cross-section that depends on the particle pathlength and may also incorporate spatial and angular dependence \cite{VasquesLarsen14}. Solution methods for the GLBE have been analyzed using Monte Carlo transport, explicit pseudo-time stepping \cite{Vasques17}, and implicit pseudo-time stepping \cite{Krycki15}. \\

The GLBE is a promising approach for more general problems, but studies thus far have been limited in scope and have yet to address problems involving finite boundaries with sources, and the problem of computing the non-classical cross-section is a potential roadblock for using this approach. In this work, the GLBE will be reviewed and extended to incorporate finite boundaries. The numerical solution using implicit time-stepping and source iteration is considered in order to leverage discrete ordinates methods for solving the resulting equations. Methods for computing the non-classical cross-section are demonstrated, and analytical approximations for binary Markovian media and low-variability Gaussian processes are derived. Finally, the GLBE solution will be shown for two problems in a square enclosure and validated against a Monte Carlo approach for a medium described by a truncated Gaussian process.\\



\section{Generalized Linear Boltzmann Equation}

As derived in \cite{LarsenVasques11}, the generalized linear Boltzmann equation (GLBE) for the angular flux $\psi(\vec{r},\hat{\Omega},s)$ at a location $\vec{r}$ with direction $\hat{\Omega}$ of particles that have traveled a distance $s$ is given by

\begin{equation}
\frac{\partial \psi}{\partial s} + \hat{\Omega}\cdot\nabla \psi + \Sigma_t(s)\psi = \delta(s)c\int_{4\pi}\int_0^\infty P(\hat{\Omega}'\cdot\hat{\Omega})\Sigma_t(s')\psi(s')ds' d\Omega' + \delta(s)\frac{Q}{4\pi}
\end{equation}

where $\Sigma_t(s)$ is the non-classical cross-section, $c$ is the single scattering fraction, $P(\hat{\Omega}'\cdot\hat{\Omega})$ is the scattering phase function, $Q$ is an arbitrary particle source term, and $\delta(s)$ denotes the Dirac delta function.\\

The GLBE may be recast into a more convenient formulation by separately considering the cases where $s>0$ and the limit of $s \to 0$. By first considering the case where $s>0$, we find the GLBE simplifies to an initial value problem:

\begin{equation}
\frac{\partial \psi}{\partial s} + \hat{\Omega}\cdot \nabla \psi + \Sigma_t(s)\psi = 0.
\label{advect}
\end{equation}

In the limit of $s \to 0$, an initial condition is derived by integrating the GLBE over the range $[-\varepsilon,\varepsilon]$ in the limit that $\varepsilon\to 0$:
\begin{equation}
\psi(\vec{r},\hat{\Omega},s=0) = c\int_{4\pi}\int_0^\infty P(\hat{\Omega}'\cdot\hat{\Omega})\Sigma_t(s')\psi(s')ds' d\Omega' + \frac{Q(\vec{r})}{4\pi}.
\label{initial}
\end{equation}

This initial condition is implicit in the case where the stochastic medium scatters particles. However, in non-scattering cases, the initial condition is fully specified. With a solution for the angular flux, the classical angular flux may be derived by integrating the angular flux over all particle path lengths

\begin{equation}
\Psi(\vec{r},\hat{\Omega}) = \int_0^\infty \psi(\vec{r},\hat{\Omega},s)ds,
\end{equation}

The particle scalar flux $\Phi^0$ and current $\Phi^1$ may be derived by taking the zeroth angular moment of the ensemble-averaged angular flux
\begin{equation}
\Psi^{(0)} = \int_{4\pi}\Psi d\Omega.
\end{equation}
\begin{equation}
\Psi^{(1)} = \int_{4\pi}\Psi \hat{\Omega}d\Omega
\end{equation}

The divergence of the particle current $\nabla \cdot \Psi^{(1)}$ may be derived from taking the moment of Equation \ref{advect}:
\begin{equation}
\int_{4\pi}\int_0^\infty\left( \frac{\partial \psi}{\partial s} +\hat{\Omega}\cdot \nabla \psi + \Sigma_t(s)\psi\right) ds d\Omega = 0
\end{equation}
Carrying out the integrals and applying Equation \ref{initial} gives a direct formula for the divergence of the current as
\begin{equation}
\nabla \cdot \Psi^{(1)} = Q(\vec{r})-(1-c)\int_{4\pi}\int_0^\infty \Sigma_t(s)\psi(s,\hat{\Omega}) ds d\Omega 
\end{equation}

\subsection{Boundary Conditions}

Previous discussions \cite{LarsenVasques11,Krycki15} of the GLBE have assumed the case of infinite domains to simplify the analysis of transport in stochastic media. As we are concerned with applying the GLBE to realistic problems, it is important to allow for finite boundaries and particle sources entering the domain from a boundary. A specified boundary source $\Psi_o$ at boundary locations $\vec{r}_w$ may be incorporated into the GLBE by applying the initial conditions

\begin{equation}
\psi(\vec{r}_w,\hat{\Omega},s) = \delta(s)\Psi_o(\vec{r}_w,\hat{\Omega})
\end{equation}

This boundary condition is analogous to the source term in the GLBE, and is applied before any particles have collided. Although this work only considers absorbing walls, a diffusely reflecting wall with reflectance $\rho$ may be modeled by
\begin{equation}
\psi(\vec{r}_w,\hat{\Omega},s=0) =\delta(s) \frac{\rho}{\pi} \int_0^\infty \int_{\hat{\Omega}\cdot\hat{n}<0} \psi(\vec{r}_w,\hat{\Omega},s)|\hat{n}\cdot\hat{\Omega}| d\Omega ds
\end{equation}
where $\hat{n}$ is the wall normal vector.

\section{Solution of the GLBE}

An efficient solution method for the GLBE will be described in this section. This method is closely related to the standard source iteration solver that is commonly applied to the steady linear Boltzmann equation. 

\subsection{Fixed Point Iteration}

Equations \ref{advect} and \ref{initial} together may be used to form an iterative solver for the GLBE. With some guess for the pre-collision angular flux $\psi^l(\vec{r},\hat{\Omega},s=0)$ at iteration $l$, the angular flux at all path lengths may be computed by solving the equation

\begin{equation}
\frac{\partial \psi_l}{\partial s} + \hat{\Omega}\cdot \nabla \psi_l + \Sigma_t(s)\psi_l = 0.
\end{equation}

The angular flux initial condition at iteration $l+1$ is computed by evaluating the following equation:

\begin{equation}
\psi_{l+1}(\vec{r},\hat{\Omega},s=0) = c\int_{4\pi}\int_0^\infty P(\hat{\Omega}'\cdot\hat{\Omega})\Sigma_t(s')\psi_l(s')ds' d\Omega' + \frac{Q}{4\pi}.
\end{equation}

The solution may be repeated until convergence is achieved. For the purposes of this work, the convergence metric is taken to be the relative norm of the residual of the classic angular flux $\epsilon^{l+1}$:

\begin{equation}
\epsilon_{l+1} = \frac{|\Psi_{l+1}-\Psi_{l}|}{|\Psi_l|}
\end{equation}

\subsection{Convergence}

The solution approach described here has the same spectral radius as the source iteration solver. To prove this, the case of classical particle transport is assumed, i.e. $\Sigma_t(s)$ is a constant for all $s$. Furthermore, we consider one-dimensional transport in the $z$-direction with angular coordinate $\mu = \hat{z}\cdot \hat{\Omega}$ in an infinite domain and isotropic scattering. In that case, the equations reduce to
\begin{equation}
\frac{\partial \psi_l}{\partial s} + \mu \frac{\partial \psi_l}{\partial z} + \Sigma_t \psi_l = 0
\end{equation}

\begin{equation}
\psi_{l+1}(z,\mu,s=0) = \frac{c}{2} \int_{-1}^1 \int_0^\infty  \Sigma_t \psi_l(s')ds' d\mu + \frac{Q}{4\pi}
\end{equation}

The system of equations may be recast in terms of the residual $f_l(z,\mu,s) = \psi(z,\mu,s)-\psi_l(z,\mu,s)$:
\begin{equation}
\frac{\partial f_l}{\partial s} + \mu \frac{\partial f_l}{\partial z} + \Sigma_t f_l = 0
\label{residual_advect}
\end{equation}

\begin{equation}
f_{l+1}(z,\mu,s=0) = \frac{c}{2}\int_{-1}^1 \int_0^\infty \Sigma_t f_l(s')ds' d\mu 
\label{residual_initial}
\end{equation}

In order to understand the convergence behavior of the iterative system, the residual function is expanded in Fourier modes with wavenumber $\lambda$ as
\begin{equation}
f_l = A \rho^l \exp(i \lambda \Sigma_t z)g(s,\mu).
\end{equation}
This expansion is similar to the method used in \cite{AdamsLarsen02} to analyze the convergence of the source iteration method. The extra step needed in this case is to determine the Fourier modes associated with the pathlength variable $s$. By applying the Fourier expansion to Equation \ref{residual_advect}, we can solve for $g(s,\mu)$:

\begin{equation}
\frac{\partial g}{\partial s} +(1+ i \mu \lambda )\Sigma_t g(s,\mu)  = 0
\end{equation}

\begin{equation}
g = h(\mu)\exp\left(-(1+i\mu \lambda)\Sigma_t s\right)
\end{equation}
where $h(\mu)$ is a function of $\mu$ resulting from the integration. The residual function is now given by
\begin{equation}
f_l = A \rho^l \exp(i \lambda \Sigma_t z)\exp\left( - (1+i\mu\lambda)\Sigma_t s\right) h(\mu) 
\end{equation}
It remains now to solve for $h(\mu)$ by using Equation \ref{residual_initial}:
\begin{equation}
\rho h(\mu) = \frac{c}{2}\int_{-1}^1 \int_0^\infty \exp\left( - (1+i\mu\lambda)\Sigma_t s' \right) h(\mu) \Sigma_t ds' d\mu
\end{equation}
Performing the integration over $s$ gives the familiar integral equation
\begin{equation}
\rho h(\mu) = \frac{c}{2} \int_{-1}^1 \frac{h(\mu)}{1+i\lambda \mu} d\mu
\end{equation}
where $h(\mu)$ is clearly a constant which may be removed from the equation. The result gives the spectral radius for each wavenumber
\begin{equation}
\rho_\lambda = \frac{c}{\lambda}\tan^{-1}\lambda
\end{equation}
with the slowest converging mode, determining the convergence of the iterative algorithm, being in the limit of $\lambda \to 0$
\begin{equation}
\rho_{max} = c
\end{equation}
This result is identical to that of the source iteration algorithm. Evidently the iterative approach for solving the GLBE retains the same convergence behavior as the standard source iteration method in the limit of classical particle transport. For non-classical transport, i.e. the case where $\Sigma_t(s)$ is variable in $s$, it is expected that one could use the maximum value of $\Sigma_t(s)$ over all $s$ as an upper bound for the convergence analysis and show that the spectral radius is still $\rho_{max}=c$.\\

The fact that the spectral radius of the iterative approach for solving the GLBE is identical to that of source iteration means that the GLBE solver inherits the same advantages and disadvantages as source iteration. The well-defined spectral radius means that one can predict the approximate number of iterations required to reach a desired level of convergence, and in the limit that the medium is purely absorbing ($c=0$), no iterations are required at all. However, in strongly scattering media ($c \to 1$), the spectral radius approaches $1$, and the iterative solver may become an impractical choice. \\

\subsection{Synthetic Acceleration}

As an improvement on source iteration, \cite{AdamsLarsen02} demonstrated that synthetic acceleration methods based on lower order forms of the transport equation may be successfully employed as preconditioners. Similar equations can be derived for the GLBE iterative solver. At iteration $l$, the following equation is solved to derive the angular flux:
\begin{equation}
\frac{\partial \psi_l}{\partial s} + \hat{\Omega}\cdot \nabla \psi_l + \Sigma_t(s) \psi_l = 0.
\end{equation}

As in the previous section, the residual between the current iterate and the true solution is defined as

\begin{equation}
f_l = \psi - \psi_l
\end{equation}

for which a transport equation may be derived

\begin{equation}
\frac{\partial f_l}{\partial s} + \hat{\Omega}\cdot \nabla f_l + \Sigma_t(s)f_l = 0,
\label{TSA}
\end{equation}

along with the initial condition

\begin{equation}
f_l(s=0) = - \psi_{l}(s=0) + c\int_{4\pi}\int_0^\infty P(\hat{\Omega'}\cdot\hat{\Omega})\Sigma_t(s')(\psi_l(s')+f_l(s'))ds'd\Omega'.
\label{TSA_init}
\end{equation}

Having solved for $f_l$, the initial condition for the following iterate is determined by

\begin{equation}
\psi_{l+1}(s=0) = c\int_{4\pi}\int_0^\infty P(\hat{\Omega'}\cdot\hat{\Omega})\Sigma_t(s')(\psi_l(s') + f_l(s'))ds'd\Omega'
\label{TSA_step}
\end{equation}

and the classical angular flux is given by
\begin{equation}
\Psi_{l+1} = \int_0^\infty f_l(s) ds + \Psi_{l}.
\end{equation}

If one were able to solve Equations \ref{TSA}, \ref{TSA_init}, and \ref{TSA_step} exactly, the true solution would be achieved in one step. However, those equations are just as difficult to solve as the original problem. As in \cite{AdamsLarsen02}, one could invoke a lower order approximation for the GLBE operator in Equation \ref{TSA}, such as transport synthetic acceleration or using the diffusion approximation derived in \cite{LarsenVasques11} for diffusion synthetic acceleration. However, we note that the integration in the pathlength variable allows for solving Equation \ref{TSA} using a coarser pathlength stepsize, and that the number of iterations required to solve Equation \ref{TSA_init} may also be limited for further efficiency gains in the inner loop. This strategy is here referred to as pathlength synthetic acceleration. Provided that the coarsening is not too aggressive, pathlength synthetic acceleration provides a convenient approach to reuse the GLBE solver to improve the rate of convergence, and may even be combined with transport synthetic acceleration.\\

\subsection{Implicit Discretization}

The problem of discretizing and solving the GLBE using the iterative solver is now considered. It is necessary to track the evolution of $\psi_l$ as a function of $s$ with high accuracy in order to evaluate the initial condition and classical angular flux accurately. At the same time, it is important to reduce the cost of the integration, as many iterations may be required to achieve a converged solution. If $\Sigma_t(s)$ is not a strong function of $s$ or $\psi_l$ is slowly changing, it is possible to use larger steps in the pathlength integration $\Delta s$ in achieving a steady solution. Thus we consider implicit trapezoidal integration in $s$ as in a pseudo-time integration, allowing for second order accurate convergence both in the evaluation of $\psi_l(s)$ as well as $\Psi_l$. An implicit discretization also has the advantage of allowing for a step size that is not limited by the spatial grid size for stability. In this work we assume the step size is fixed without loss of generality. The resulting semi-discrete equation for the angular flux is

\begin{equation}
\frac{\psi_l^{n+1}-\psi_l^{n}}{\Delta s} + \frac{1}{2}\left(\hat{\Omega}\cdot \nabla \psi_l^{n+1} +  \hat{\Omega}\cdot \nabla \psi_l^{n}\right) + \frac{1}{2}\left(\Sigma_t(s^{n+1})\psi_l^{n+1}+\Sigma_t(s^n)\psi_l^n\right) = 0
\end{equation}

where $\psi_l^n$ and $s^n$ are the angular flux and pathlength at step $n$ respectively. By regrouping terms for the angular flux at $n+1$, the discretized equation is

\begin{equation}
\hat{\Omega}\cdot \nabla \psi_l^{n+1} + \left( \Sigma_t(s^{n+1})+ \frac{2}{\Delta s}\right)\psi_l^{n+1} = -\hat{\Omega}\cdot \nabla \psi_l^{n} + \left( -\Sigma_t(s^{n})+ \frac{2}{\Delta s}\right)\psi_l^{n},
\end{equation}

which has the same form as the quasi-steady Boltzmann transport equation with pseudo-time dependent cross-section and source terms. By applying an angular discretization via the discrete ordinates method and combining it with a spatial discretization such as diamond differencing or discontinuous Galerkin methods, the resulting equation may be efficiently inverted for $\psi_l^{n+1}$ by using full domain sweeps.\\

With a trapezoidal integration in $s$, the classical angular flux and initial condition are given by
\begin{equation}
\Psi_{l} = \frac{\Delta s}{2}\left( \psi_l^0 + \psi_l^N\right) + \sum_{n=1}^{N-1} \psi_l^n \Delta s
\end{equation}

\begin{equation}
\psi_{l+1}(s=0) = \frac{c}{4\pi}\int_{4\pi}\left(\frac{\Delta s}{2}\left( \Sigma_t(s^0)\psi_l^0 + \Sigma_t(s^N)\psi_l^N\right) + \sum_{n=1}^{N-1}\Delta s \Sigma_t(s^n) \psi_l^n\right)d\Omega + \frac{Q}{4\pi}
\end{equation}

Rather than storing $\psi_l$ for all $s$, the classical angular flux and initial condition may be updated incrementally as the integration proceeds to reduce computational memory requirements.\\

The final consideration that needs to be made is for the boundary condition. A simple approach is used here to regularize the Dirac delta function in the boundary condition by approximating it as a rectangular pulse of duration $\Delta s$ and height $1/\Delta s$, leading to

\begin{equation}
\psi(\vec{r}_w,\hat{\Omega},s=0) \approx \frac{\Psi_o(\vec{r}_w,\hat{\Omega})}{\Delta s}.
\end{equation}

\section{Application to Transport in Two-Dimensional Stochastic Media}

The solution of the GLBE will be demonstrated in this section for a few different example problems, and the results will be compared against Monte Carlo runs to demonstrate convergence of the GLBE solution. A process for generating the cross-section $\Sigma_t(s)$ is described first. In this section we will also make the simplifying assumption of isotropic scattering, i.e. $P(\hat{\Omega}'\cdot \hat{\Omega}) = 1/4\pi$ without loss of generality.\\

\subsection{Generation of Stochastic Cross Sections}

In this work the case of stochastic media determined by Gaussian processes with a spatially constant mean $\Sigma_o$ and a Gaussian covariance function is considered. Then independent realizations of stochastic media may be constructed by using Karhunen-Loeve expansions as in \cite{Fichtl11}. Starting from the Gaussian covariance function with variance $\sigma^2$ and correlation length $l$

\begin{equation}
C(\vec{r},\vec{r}') = \sigma^2 \exp\left(-\frac{|\vec{r}-\vec{r}'|^2}{2l^2}\right),
\end{equation}

the covariance matrix may be evaluated on a discrete grid. With a covariance matrix of size $M \times M$, the $M$ eigenvalues $\lambda_i$ and eigenvectors $v_i$ are computed from

\begin{equation}
Cv_i = \lambda_i v_i.
\end{equation}

Defining $Z_i$ as independent Gaussian random variables with zero mean and unit variance, realizations of the cross-section are generated as

\begin{equation}
\Sigma_t(\vec{r},\xi) = \Sigma_o + \sum_{i=1}^M \sqrt{\lambda_i}Z_i v_i(\vec{r})
\end{equation}

Since this process may lead to negative values of $\Sigma_t$, negative values are truncated to zero. This results in an increase in the mean and decrease in the variance.\\


Once many realizations of $\Sigma_t$ have been generated, the pathlength distribution and non-classical cross section may be computed. Following the definition of the pathlength probability distribution function in \cite{LarsenVasques11}

\begin{equation}
p(s) =  \left\langle \Sigma_t(s)\exp\left( -\int_0^s \Sigma_t(s')ds'\right)\right\rangle
\end{equation}

with the brackets indicating an ensemble-average, many one-dimensional stochastic realizations of $\Sigma_t(s)$ are generated, and the integral is computed numerically using trapezoidal integration. With a well-resolved $p(s)$ in hand, the non-classical cross-section is computed from

\begin{equation}
\Sigma_t(s) = \frac{p(s)}{1-\int_0^{s}p(s')ds'}
\end{equation}

with interpolation to a finer grid as needed for increased accuracy in the GLBE solution. Care is also taken to ensure the length of the domain is long enough that a very small number of particles survive to the end; without this measure, the above equation may yield unphysical results for the cross-section.\\

\begin{figure}
\centering
\includegraphics[scale=.4]{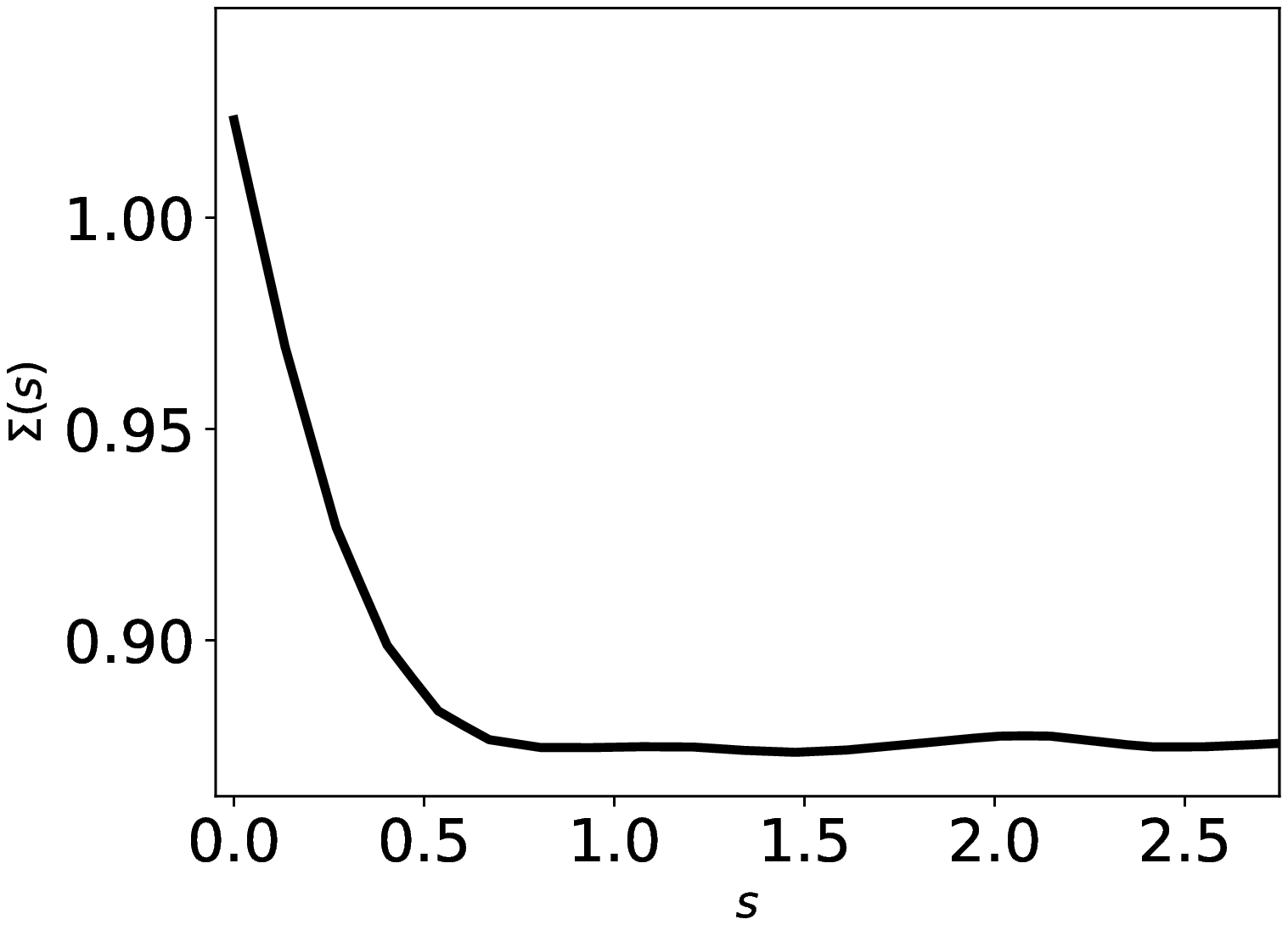}
\includegraphics[scale=.4]{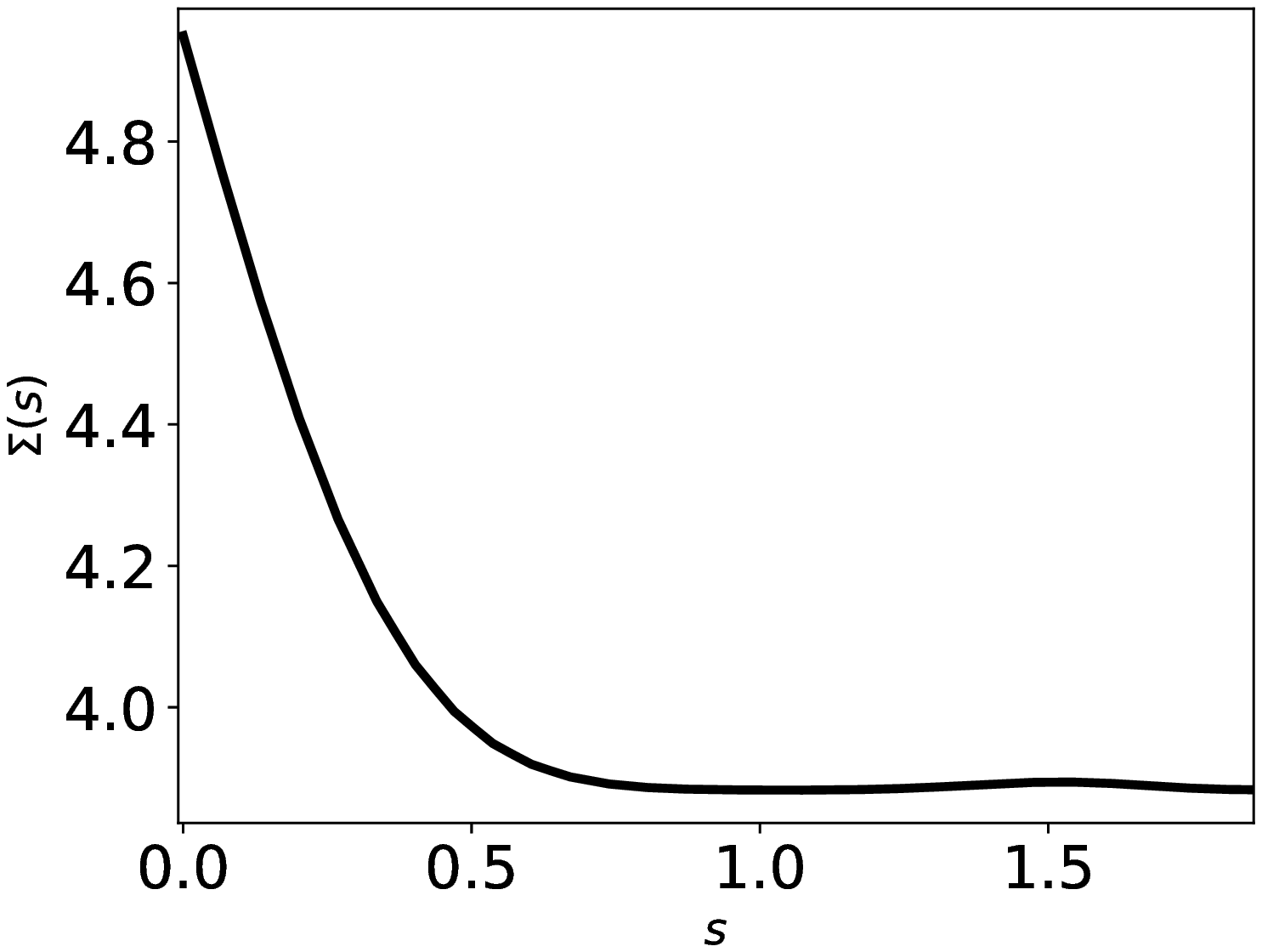}
\includegraphics[scale=.4]{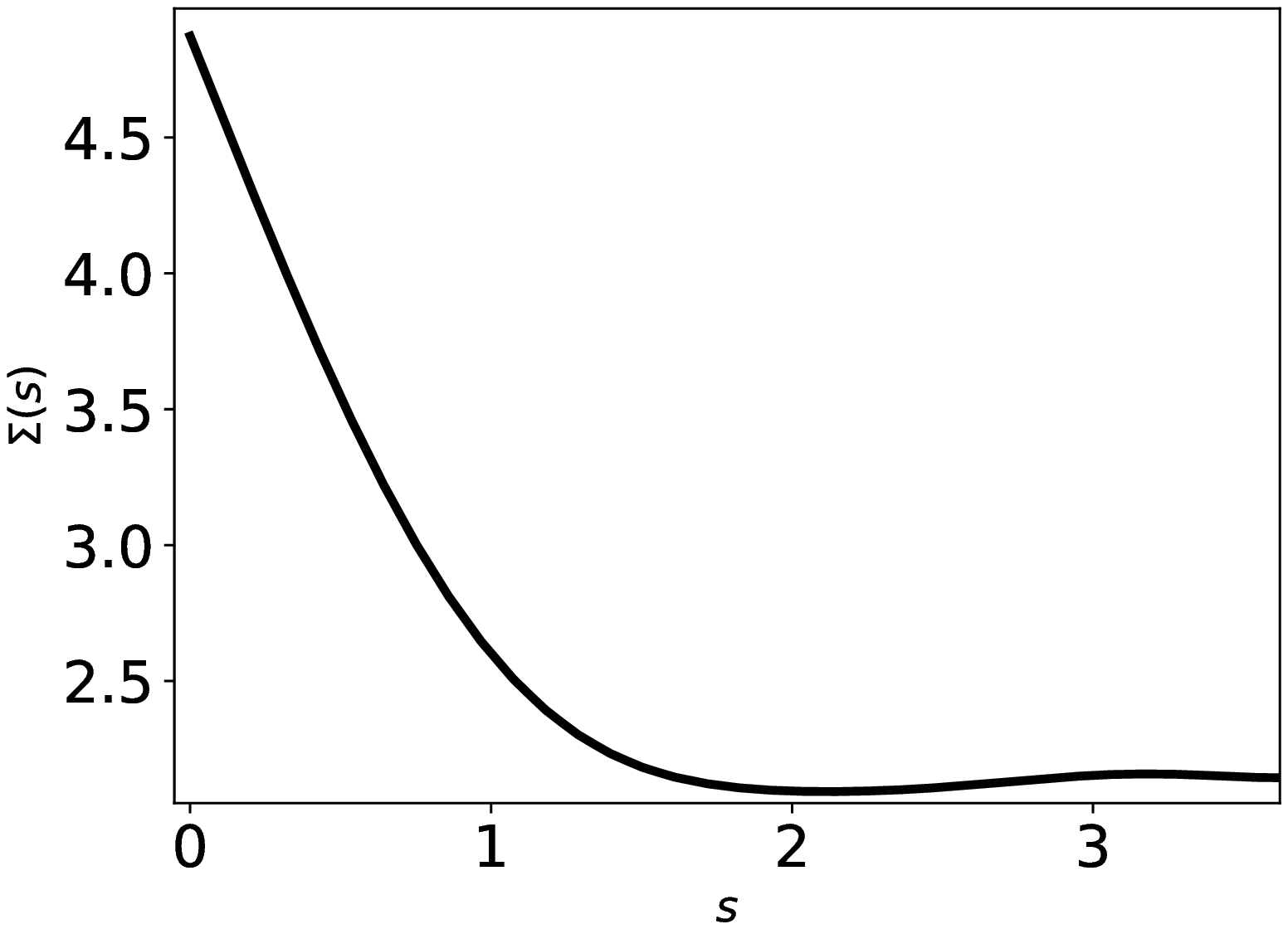}
\caption{The non-classical cross-section for truncated Gaussian process media with parameters given by $(\mu,\sigma^2,l)$: a) (1,0.5,0.3), b) (5,3,0.3), c) (5,3,1).}
\label{cross_sect}
\end{figure}

Figure \ref{cross_sect} shows a few examples of the non-classical cross-section for different truncated Gaussian processes. The resulting curves reflect the underlying statistics of the problem. Initially, all of the particles enter into the medium with the same rate of attenuation on average. Particles in regions of high material density tend to be extinguished faster, so the effective cross-section that the remaining particles see decreases. Past a distance on the order of the correlation length of the medium, the attenuation rate equilibrates as particles see a mixture of high and low density material, and the cross-section asymptotes to a constant cross-section identical to a homogenized constant. The figures also represent differing magnitudes of variation in the cross-section, with lower coefficients-of-variation corresponding to smaller decreases, and longer correlation lengths corresponding to longer decay lengths.

\subsection{Analytical approximations}

In some cases, it may not be feasible to estimate the non-classical cross-section directly from Monte Carlo sampling; the number of realizations required, along with the spatial resolution and maximum pathlength required for the analysis, may be prohibitively expensive. It is useful to have convenient approximations available in that case. In other cases, the mean line transmission $\langle \Psi(s)\rangle$ may be available already, so $\Sigma_t(s)$ may be estimated directly from
\begin{equation}
\Sigma_t(s) = -\frac{1}{\langle \Psi(s)\rangle}\frac{d\langle \Psi(s)\rangle}{ds}
\end{equation}

Analytical non-classical cross-sections are derived in the following sections for the special cases of binary Markovian media and Gaussian processes with low coefficient of variation.

\subsubsection{Binary Markovian media}

For discrete mixtures of two separate components, the approximation of a binary Markovian medium is commonly made. This approximation assumes that the pathlength of a ray passing though any one component of the mixture has an exponential distribution. Using the results of \cite{LevermorePomraning}, the mean line transmission for a binary Markovian medium is given by
\begin{equation}
\langle \Psi(s)\rangle = \left( \frac{r_+ - \tilde{\sigma}}{r_+ - r_-}\right) \exp(-r_+ s) + \left( \frac{\tilde{\sigma}-r_-}{r_+ - r_-}\right) \exp(-r_- s)
\end{equation}

where the various parameters $r_\pm$, $\tilde{\sigma}$, and $\langle \sigma \rangle$ are given by

\begin{equation}
r_\pm = \frac{1}{2}(\langle \sigma \rangle + \tilde{\sigma} \pm \sqrt{(\langle \sigma \rangle - \tilde{\sigma})^2 + 4\beta})
\end{equation}

\begin{equation}
\tilde{\sigma} = p_1 \sigma_1 + p_2 \sigma_2 + \frac{1}{\lambda_1} + \frac{1}{\lambda_2}
\end{equation}

\begin{equation}
\beta = (\sigma_1-\sigma_2)^2 p_1 p_2
\end{equation}

\begin{equation}
\langle \sigma \rangle = p_1 \sigma_1 + p_2 \sigma_2
\end{equation}

where $\sigma_1$ and $\sigma_2$ are the cross-sections for the two components in the mixture, $p_1$ and $p_2$ are the volume fractions of each component, and $\lambda_1$ and $\lambda_2$ are the average path-lengths within each fluid.\\

The choice of a Markovian model for the mixture is convenient due to the simplicity of the Levermore-Pomraning model. However, this model is known to suffer from a loss of accuracy at higher scattering ratios due to the need for an additional closure condition. Instead, the above result may be used to derive the non-classical cross section for use in the GLBE as

\begin{equation}
\Sigma(s) = \frac{\left( \frac{r_+ - \tilde{\sigma}}{r_+-r_-}\right)r_+ \exp(-r_+s)
+ \left( \frac{\tilde{\sigma}-r_-}{r_+-r_-}\right)r_-\exp(r_- s) } 
{\left( \frac{r_+ - \tilde{\sigma}}{r_+ - r_-}\right) \exp(-r_+ s) + \left( \frac{\tilde{\sigma}-r_-}{r_+ - r_-}\right) \exp(-r_- s)}
\end{equation}

This analytical form would allow for convenient incorporation of a binary Markovian medium into the GLBE framework without a loss of accuracy at higher scattering ratios.

\subsubsection{Quasi-Gaussian media}

For continuously varying media with low coefficient of variation, \cite{Banko18} has demonstrated that another analytical approximation may be made, the derivation of which will be repeated here. It is assumed that the mean, variance, and two-point correlation function of the medium are known.\\

Define $\tau$ to be the optical depth across some path length in the random medium with cross-section $\sigma(x)$:
\begin{equation}
\tau = -\log(\Psi(s)/\Psi_o) = \int_0^s \sigma(x) dx
\end{equation}
Since $\sigma$ is a random field, $\tau$ is also a random variable. The mean of $\tau$ is given by
\begin{equation}
\langle \tau \rangle = \langle \sigma \rangle s
\end{equation}
Taking the decomposition of the field into a stationary mean and a fluctuating component $\sigma = \langle \sigma \rangle + \sigma'$, the fluctuation of $\tau$ is
\begin{equation}
\tau' = \int_0^s \sigma'(x) dx
\end{equation}
from which the variance may be computed as
\begin{equation}
\langle \tau'^2\rangle = \int_0^s \int_0^s \langle \sigma'(x)\sigma'(y)\rangle dx dy
\end{equation}
With some manipulation of this integral, we arrive at a convenient expression for the variance as
\begin{equation}
\langle \tau'^2 \rangle = 2\int_0^s \int_0^{s-s'} \Phi_{\sigma\sigma}(r)dr ds'
\end{equation}
which may be further simplified to
\begin{equation}
\langle \tau'^2 \rangle = 2\int_0^s (s-r)\Phi_{\sigma\sigma}(r)dr
\end{equation}
where $\Phi_{\sigma\sigma}(r) = \langle \sigma'(z)\sigma'(z+r)$ is the stationary two-point correlation function, independent of location $z$. \\

With the approximation that $\tau$ has a nearly normal distribution (in which we neglect the probability that $\tau$ is negative), $\Psi(s)/\Psi_o$ is log-normal distributed. Using the properties of a log-normal distribution, the mean transmission is given by
\begin{equation}
\langle \Psi(s)/\Psi_o \rangle = \exp\left(-\langle \sigma \rangle s +\int_0^s (s-r)\Phi_{\sigma\sigma}(r)dr ds'\right)
\end{equation}
from which we compute the relevant non-classical cross section as
\begin{equation}
\Sigma(s) = \langle\sigma\rangle - \int_0^{s}\Phi_{\sigma\sigma}(r)dr
\end{equation}

For any sufficiently compact two-point correlation, the non-classical cross-section is initially equal to the atomic-mixing approximation and asymptotes to a reduced constant value (for a positively correlated medium). This reduction accounts for the effects of correlation in the medium; the particles that stream initially through high density regions of the medium tend to be absorbed and the average cross-section seen by the surviving particles tends to decrease. This asymptotic value was derived in \cite{Frankel17,FrankelDissertation} for the case where the two-point correlation was approximately a Dirac delta function or $s \to \infty$. This result is far more general for arbitrary functions with the caveat that $\tau$ has a non-zero probability of being negative. This connection suggests that the full non-classical cross-section for various distributions of media may be used to derive homogenization models for lower fidelity modeling. Alternatively, \cite{Banko18} derived an alternative model for transmission statistics that guarantees strict positivity in the non-classical cross-section. Such a model may prove to be more accurate in future work, though it is analytically complex.

\begin{figure}
\centering
\includegraphics[scale=.5]{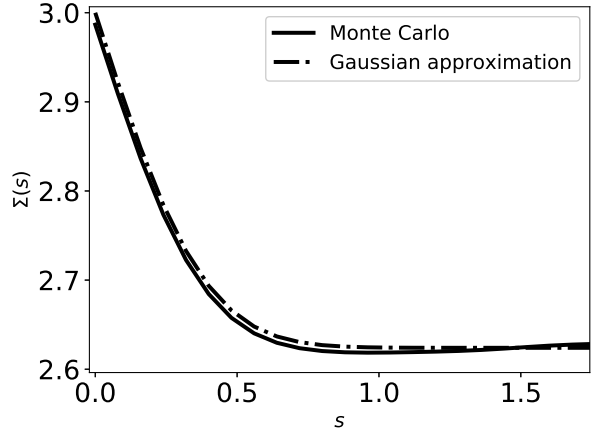}
\caption{Comparison between the directly computed non-classical cross-section and the Gaussian approximation for a truncated Gaussian process medium with $\mu=3$, $\sigma^2=1$, and $l=0.3$.}
\label{gauss_approx}
\end{figure}

\section{Example Problems}

In this section, the solution of the GLBE is benchmarked against direct sampling of the stochastic medium solved with the discrete ordinates method. The atomic mixing solution is also shown for reference where the cross-section is the mean of the truncated Gaussian.\\

In the examples below, level symmetric quadrature was used for the angular discretization, and diamond differencing was used for the spatial discretization without a negative-flux fixup. The domain was taken to be a square of length $1$ on each side discretized into a uniform 20x20 grid, and the walls were purely absorbing. The time advancement was stopped when the increments in the integrated angular flux reached below $10^{-7}$, and the outer scattering loop was stopped when the residual in the angular flux reached below $10^{-6}$.\\

\subsection{Uniform source}

\begin{figure}
\centering
\includegraphics[scale=.5]{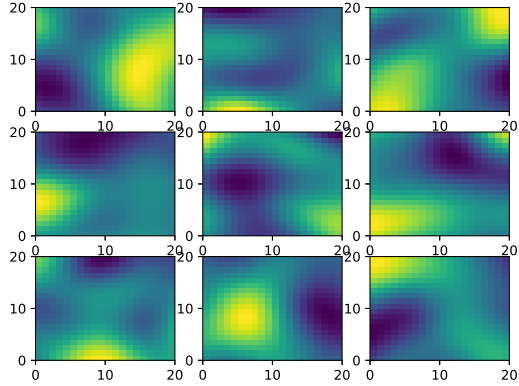}
\caption{Example cross-section fields for the uniform source problem.}
\label{example_fields_source}
\end{figure}

In this problem, there are no external sources, and a uniformly distributed source of magnitude $1$ is prescribed in the domain. The medium is a truncated Gaussian process with nominal parameters $\mu=5$, $\sigma^2 = 2$, and $l=0.3$ with a stepsize of $\Delta s = 0.008$. A set of realizations of this process are shown in Figure \ref{example_fields_source}. The $S_8$ level symmetric quadrature set was used.\\

\begin{figure}
\centering
\includegraphics[scale=.5]{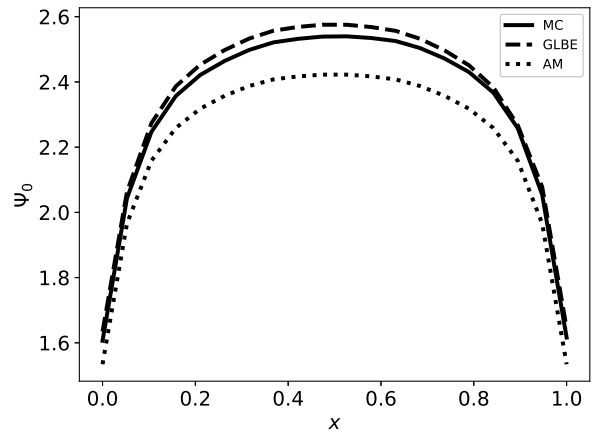}
\caption{The integrated scalar flux along $x=0.5$ for the GLBE, the Monte Carlo samples (MC), and the atomic mixing reference solution (AM) for no scattering with a uniform, isotropic source.}
\label{source_noscat}
\end{figure}

\begin{figure}
\centering
\includegraphics[scale=.5]{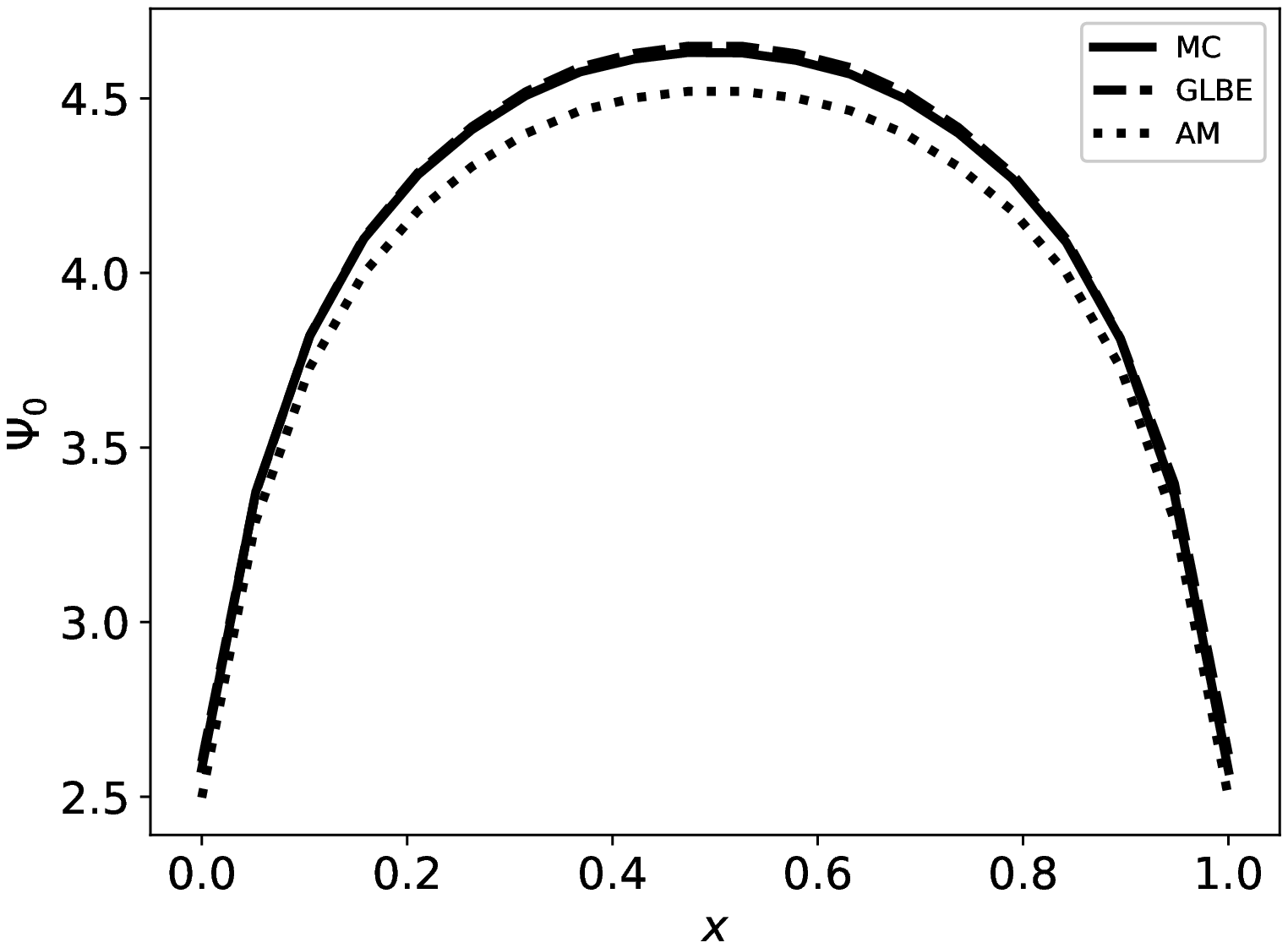}
\caption{The integrated scalar flux along $x=0.5$ for the GLBE, the Monte Carlo samples (MC), and the atomic mixing reference solution (AM) for a scattering ratio of $0.5$ with a uniform, isotropic source.}
\label{source_scat}
\end{figure}

Figure \ref{source_noscat} shows the integrated scalar flux along the centerline of the domain for the medium without any scattering, comparing the GLBE solution with Monte Carlo sampling and the atomic mix approximation. The GLBE solution agrees closely with the Monte Carlo results, with some discrepancy in the middle of the domain. It is likely that higher resolution and more accurate discretizations would narrow the gap in the solutions. Figure \ref{source_scat} supports this, which shows the results for the same problem with a scattering ratio of $c=0.5$. The difference in the results in this case is very small, and the solutions are nearly identical. However, just over $6000$ sweeps were necessary to complete the solution of the GLBE compared to over $17000$ sweeps in the aggregated Monte Carlo samples, making the GLBE solution substantially more attractive for the same degree of accuracy.\\

\subsection{Emitting surface}

\begin{figure}
\centering
\includegraphics[scale=.5]{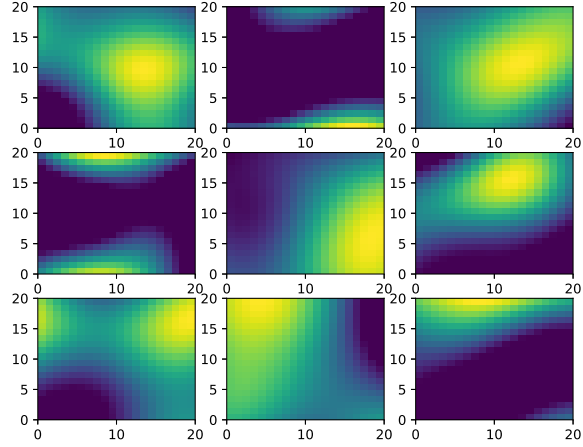}
\caption{Example cross-section realizations for the emitting surface problem.}
\label{example_fields_wall}
\end{figure}

In this problem, an isotropically emitting source of magnitude $1$ is affixed to one of the surfaces in the domain, and no internal sources are present. The medium is a truncated Gaussian process with nominal parameters $\mu = 1$, $\sigma^2 = 2$, and $l=0.5$ with a stepsize of $\Delta s = 0.007$. Example fields for this problem are shown in Figure \ref{example_fields_wall}. The $S_{16}$ level symmetric quadrature set was used.\\

\begin{figure}
\centering
\includegraphics[scale=.5]{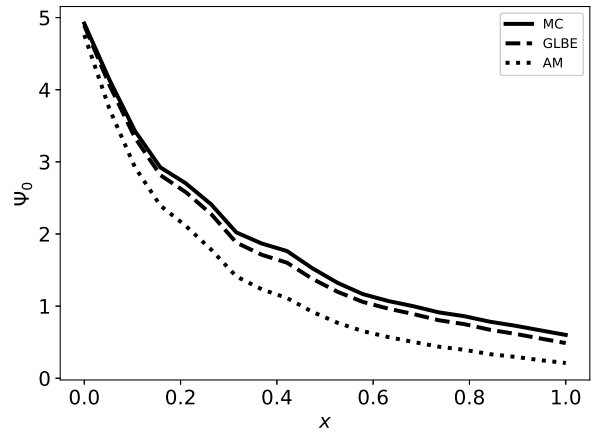}
\caption{The integrated scalar flux along $x=0.5$ for the GLBE, Monte Carlo samples, and the atomic mixing reference solution for no scattering with a boundary source.}
\label{wall_noscat}
\end{figure}

\begin{figure}
\centering
\includegraphics[scale=.5]{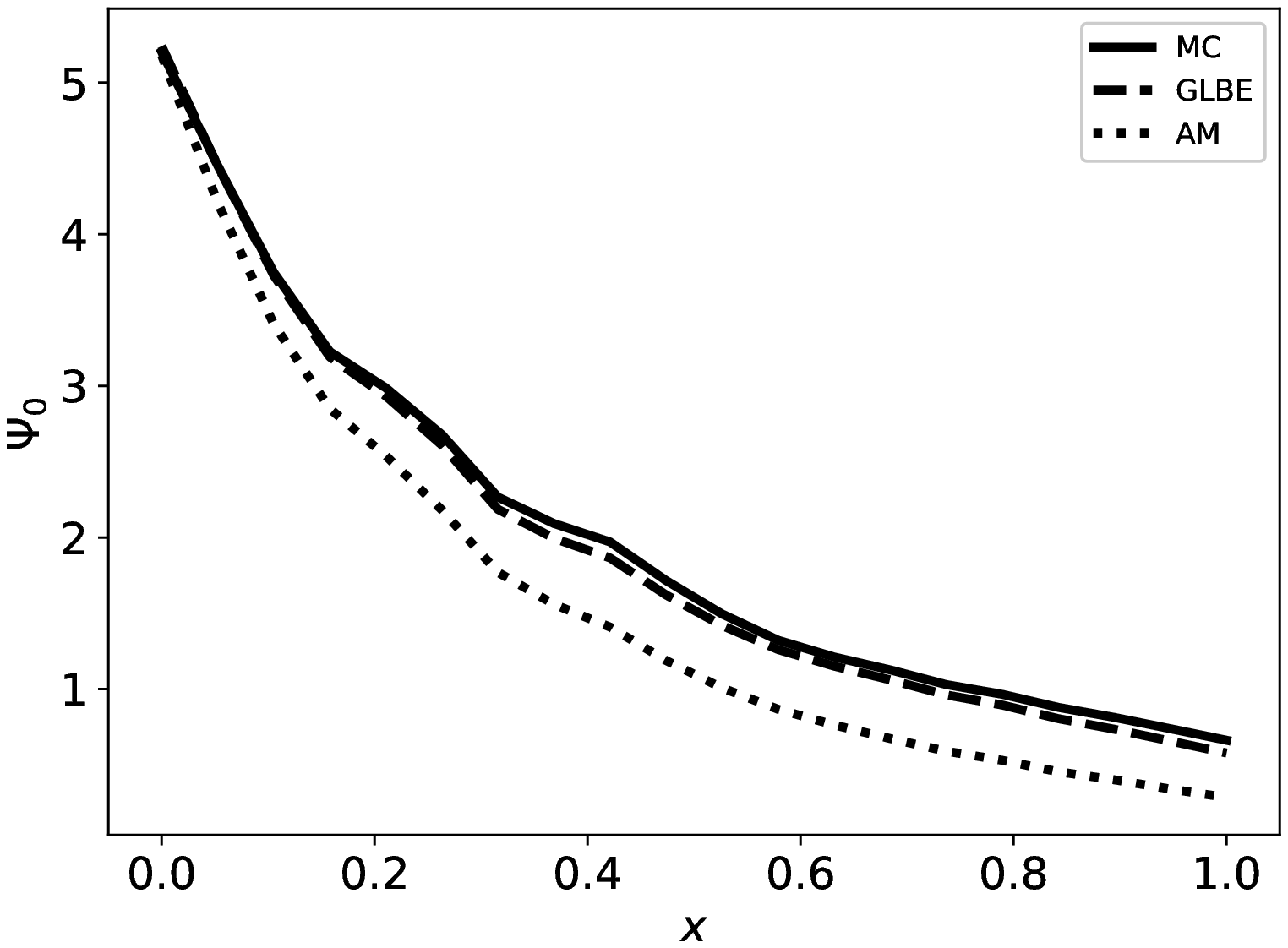}
\caption{The integrated scalar flux along $x=0.5$ for the GLBE, Monte Carlo samples, and the atomic mixing reference solution for a scattering ratio of $0.3$ with a boundary source.}
\label{wall_scat}
\end{figure}

Figure \ref{wall_noscat} shows the integrated scalar flux for the emitting surface along the axis normal to the wall. Although some discretization artifacts are present, the results from the Monte Carlo sampling and GLBE solution are nearly identical, and show substantially higher transmission through the domain than the atomic mixing model predicts. Figure \ref{wall_scat} shows the same problem for the case where $c=0.3$. In the scattering case, the aggregated number of GLBE sweeps was over $5000$, whereas over $11000$ were required for the Monte Carlo samples.\\

\subsection{Acceleration}

The pathlength synthetic acceleration strategy was tested in a problem involving a scattering ratio of $0.9$ in a truncated Gaussian process medium of mean $10$, variance $5$, and correlation length $0.5$. This setting is expected to be challenging for a source iteration strategy due to the high optical depth and scattering ratio. The high-resolution cross-section was evaluated with a pathlength stepsize of $\Delta s = 0.00214$, and the low-resolution was evaluated with a fivefold coarsening of $\Delta s = 0.0107$. Only one inner iteration was performed with the low-resolution cross-section. The problem considered was the uniform source problem described above.\\

\begin{figure}
\centering
\includegraphics[scale=.5]{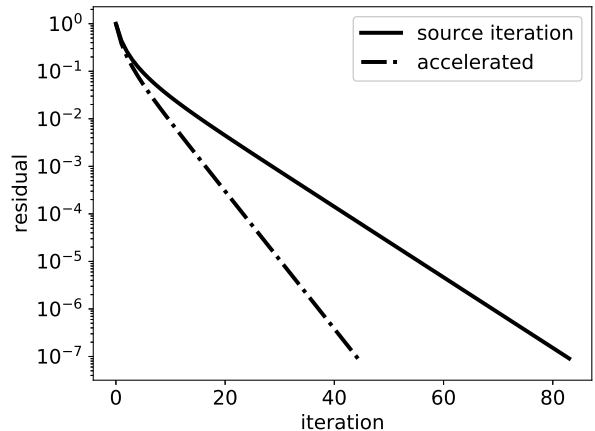}
\caption{Residual history for the uniform source problem with a higher optical depth and scattering ratio, comparing the source iteration and accelerated solvers.}
\label{residual_accel}
\end{figure}

The residual history is shown in Figure \ref{residual_accel} for the source iteration and accelerated solvers. The initial residuals for the two are nearly identical, but the asymptotic convergence rate of the accelerated solver is nearly half that of the source iteration solver. This is due to the inner iteration adding a substantial error reduction at each step, effectively acting as an additional outer iteration. These results are promising, but likely the pathlength synthetic acceleration would be best combined with a transport synthetic acceleration or geometric multigrid approach to maximize the performance of the solver.

\section{Conclusion}

This work has demonstrated the applicability of the GLBE to arbitrary stochastic processes in bounded multidimensional problems, and that it may be solved efficiently using existing deterministic methods. The choice of implicit discretization and source iterations lends itself to discrete ordinates methods. The non-classical cross-section may be computed directly from Monte Carlo statistics or from analytical approaches if the mean transmission is known. The resulting numerical method was tested against Monte Carlo sampling of truncated Gaussian process media in a square enclosure and showed excellent agreement in cases with and without scattering. A strategy for accelerating the convergence of the iterative process was also demonstrated to improve the rate of convergence for a particular case with a high scattering ratio.\\

Although the example problems only considered two-dimensional media, it is easy to extend this approach to three-dimensional media and unstructured grids, lending itself to immediate use in more practical problems of interest. In order to fully open the GLBE to the application space of interest, it is necessary to develop and demonstrate versions that capture charged-particle transport and reactor criticality computations. In addition, the larger scale of these problems requires that more efficient implementations of the GLBE solver be available, including versions that implement adaptive time integration, synthetic accleration or preconditioning methods, and parallel sweeps. Examining the relative effectiveness of these methods is important and should be studied further.\\

The complexity of the physics scenarios of interest is also important to consider. While this study and others before it have considered homogeneous media and uniform sources, it is important to remember that many processes are inhomogeneous and anisotropic, and that problems of interest have source terms that are dependent on the stochastic process as well (such as blackbody radiation or fission neutrons). The question of how effective the GLBE solver presented here is for such problems needs to be determined for these cases as well.\\

\section*{Acknowledgments}

Sandia National Laboratories is a multimission laboratory managed and operated by National Technology \& Engineering Solutions of Sandia, LLC, a wholle owned subsidiary of Honeywell International Inc., for the U.S. Department of Energy's National Nuclear Security Administration under contract DE-NA0003525. This paper describes objective technical results and analysis. Any subjective views or opinions that might be expressed in the paper do not necessarily represent the views of the U.S. Department of Energy of the United States Government. The author would also wish to thank Dr. Arlyn Antolak for his help and advice throughout the progress of this work, as well as the LDRD office at Sandia National Laboratories for supporting this work under project number 212586. This document is set for unclassified unlimited release under SAND2019-0528J.

\bibliographystyle{unsrt}
\bibliography{paper}

\end{document}